\documentstyle[12pt]{article}

\def\be{\begin{equation}}
\def\ee{\end{equation}}
\def\bea{\begin{eqnarray}}
\def\eea{\end{eqnarray}}

\def\bar{\overline}

\def\a{\alpha}

\def\l{\lambda}
\def\bc{\begin{center}}
\def\ec{\end{center}}
\def\O{{\cal O}}
\def\sdpro{\begin{picture}(1,10)
\put(2,0){\line(0,1){6}}
\end{picture}
\!\! \times }

\def\PR#1#2#3{Phys. Rev.  {\bf #1}, (#3) #2}
\def\PRL#1#2#3{Phys. Rev. Lett. {\bf #1}, (#3) #2}
\def\PL#1#2#3{Phys. Lett. {\bf #1}, (#3) #2}

\def\NP#1#2#3{Nucl. Phys. {\bf #1}, (#3) #2}

\def\PTP#1#2#3{Prog. Theor. Phys. {\bf #1}, (#3) #2}

\begin{document} 

\begin{flushright}
hep-ph/0107267 \ AUE-01-01 / KGKU-01-01 / MIE-01-01 
\end{flushright}

\vspace{3mm}

\begin{center}
{\large \bf Flavor Symmetry on Non-Commutative \\
       Compact Space and $SU(6) \times SU(2)_R$ Model }

\vspace{5mm}

Yoshikazu ABE 
            \footnote{E-mail address: abechan@phen.mie-u.ac.jp}, 
Chuichiro HATTORI$^a$ 
            \footnote{E-mail address: hattori@ge.aitech.ac.jp}, 
Masato ITO$^b$ 
            \footnote{E-mail address: mito@eken.phys.nagoya-u.ac.jp}, \\
Masahisa MATSUDA$^c$
            \footnote{E-mail address: mmatsuda@auecc.aichi-edu.ac.jp},
Mamoru MATSUNAGA 
            \footnote{E-mail address: matsuna@phen.mie-u.ac.jp}
and Takeo MATSUOKA$^d$ 
            \footnote{E-mail address: matsuoka@kogakkan-u.ac.jp}

\end{center}

\begin{center}
{\it 
Department of Physics Engineering, Mie University, Tsu, JAPAN 514-8507 \\
{}$^a$Science Division, General Education, Aichi Institute of 
Technology, Toyota, JAPAN 470-0392 \\
{}$^b$Department of Physics, Nagoya University, Nagoya, 
JAPAN 464-8602 \\
{}$^c$Department of Physics and Astronomy, Aichi University 
of Education, Kariya, Aichi, JAPAN 448-8542 \\
{}$^d$Kogakkan University, Nabari, JAPAN 518-0498 
}
\end{center}

\vspace{3mm}

\begin{abstract}
In the four-dimensional effective theory from string 
compactification discrete flavor symmetries arise from 
symmetric structure of the compactified space and 
generally contain both the $R$ symmetry and non-$R$ symmetry. 
We point out that a new type of non-Abelian flavor symmetry 
can also appear if the compact space is non-commutative. 
Introducing the dihedral group $D_4$ as such a new type of 
flavor symmetry together with the $R$ symmetry and 
non-$R$ symmetry in $SU(6) \times SU(2)_R$ model, 
we explain not only fermion mass hierarchies but also 
hierarchical energy scales including 
the breaking scale of the GUT-type gauge symmetry, 
intermediate Majorana masses of R-handed neutrinos and 
the scale of $\mu$-term. 
\end{abstract}

\newpage 
%%%%%%  SECTION  1  %%%%%%%%%%%%%%%%%%%%%%%%%%%%%%%%%%%%%%%%
\section{Introduction}

Quark and lepton masses and mixing angles exhibit apparent 
characteristic patterns. 
Many authors have made attempts to explain 
these characteristic patterns by introducing an appropriate 
flavor symmetry and by relying on the Froggatt-Nielsen 
mechanism\cite{F-N}. 
As for the flavor symmetry attention has been confined to 
continuous symmetry such as the gauged $U(1)$. 
However, in the string theory, which is the only known 
candidate of the unified theory including gravity, 
discrete symmetries very likely arise as 
the flavor symmetry from the symmetric structure of 
the compactified space. 
As seen from the geometrical construction of Calabi-Yau space 
and also from the algebraic construction given 
by the Gepner model\cite{Gepner}, 
in the string theory it is likely that we have both $R$ 
and non-$R$ discrete symmetries as the flavor symmetry.

On the other hand, there are several characteristic scales 
in energy regions ranging from the string scale 
$M_S$($O(10^{18})$GeV) to the electroweak scale 
$M_W$($O(10^2)$GeV). 
First, the breaking scale $M_{GUT}$ of the GUT-type gauge symmetry 
should be larger than $O(10^{16})$GeV to guarantee 
the longevity of the proton. 
Second, the seesaw mechanism for neutrinos implies that 
the Majorana mass scale $M_R$ of R-handed neutrinos is 
expected to be $O(10^{10 \sim 12})$GeV. 
Third, in order for the effective theory to be consistent with 
the standard model, 
the scale of the $\mu$-term is required to be 
$O(10^{2 \sim 3})$GeV.

In this paper we concentrate our attension on whether or not 
these characteristic scale hierarchies are derivable from 
the flavor symmetry. 
In our previous work
\cite{Matsu1}, 
in which we chose a discrete Abelian $R$ symmetry 
as the flavor symmetry together with $SU(6) \times SU(2)_R$ 
gauge symmetry, 
it was shown that the gauge symmetry is spontaneously broken 
at tree level down to the standard model gauge group 
$G_{SM} = SU(3)_c \times SU(2)_L \times U(1)_Y$ in two steps. 
Furthermore, we explained the triplet-doublet splitting as well as 
fermion mass hierarchies except for neutrinos. 
However, we could not derive the correct scale 
of the Majorana mass $M_R$ of R-handed neutrinos. 
In fact, in the previous paper\cite{Matsu1} 
we obtained the result that $M_R$ is either around colored 
Higgs mass scale or around the scale of the $\mu$-term. 
Phenomenologically, the scale $M_R$ is required to be almost 
equal to the geometrically averaged value between $M_S$ and $M_W$. 
In order for us to explain such a scale of $M_R$, 
it seemed that we need an additional flavor symmetry.

Recently, extensive studies of string with discrete torsion 
have been made\cite{torsion,torsion2}. 
The discrete torsion is inherently related to the introduction of 
nontrivial background $B$ field.
In the presence of background $B$ field coordinates of D-branes 
become non-commutative operators\cite{B-f}. 
The non-commutativity of coordinates is closely linked to 
quantum fluctuation of compactified space. 
Due to the quantum fluctuation of the compact space 
coordinates are described in terms of a projective 
representation of the discrete symmetry group. 
In the string theory massless matter fields correspond to 
the degree of freedom of the deformation of compact space. 
Various types of the deformation are represented by 
approriate functions of the coordinates with definite charges 
under the discrete group. 
Therefore, massless matter fields are also described in terms 
of the projective representation. 
On the other hand, four-dimensional effective Lagrangian of 
the theory should belong to the center of 
the non-commutative algebra, 
which is the subalgebra consisting of commuting elements of 
the algebra. 
This means, as we shall see in the following section, 
that a new type of flavor symmetry also arises 
in the compact space with non-commutative geometry.

Motivated by phenomenological requirements, in this paper 
we introduce as the flavor symmetries a discrete non-Abelian 
symmetry as well as discrete Abelian symmetries. 
As for Abelian symmetries we choose $R$-parity and 
${\bf Z}_3 \times {\bf Z}_5$ $R$ symmetry in addition to 
${\bf Z}_2 \times {\bf Z}_7$ non-$R$ symmetry. 
Further, we introduce the dihedral group $D_4$ as a 
non-Abelian symmetry and take its projective representation 
arising from the non-commutativity of coordinates. 
Applying this new type of the flavor symmetry to the 
$SU(6) \times SU(2)_R$ model, 
we explain not only fermion mass hierarchies but also 
hierarchical energy scales including the breaking scale of 
the GUT-type gauge symmetry, intermediate Majorana masses of 
R-handed neutrinos and the scale of $\mu$-term.

This paper is organized as follows. 
In section 2 it is pointed out that the string theory naturally 
provides a discrete symmetry, 
in which non-Abelian symmetry is likely contained together 
with Abelian $R$ and non-$R$ symmetries. 
The discrete symmetry plays a role of the flavor symmetry. 
When coordinates are described in terms of a projective 
representation of the discrete symmetry, 
matter fields also have matrix-valued charges of 
the discrete symmetry. 
In section 3 we briefly review the minimal $SU(6) \times SU(2)_R$ 
model and introduce Abelian $R$ and non-$R$ symmetries as 
the flavor symmetry. 
We attempt to explain not only fermion mass hierarchies but 
also hierarchical energy scales. 
It is found that an additional selection rule is needed to 
explain the hierarchical energy scales. 
In section 4 we introduce a non-Abelian discrete symmetry 
to yield the additional selection rule and 
take its projective representation. 
This non-Abelian group turns out to be just the dihedral 
group $D_4$. 
By using the flavor symmetry including $D_4$ together with 
Abelian $R$ and non-$R$ symmetries, 
mass spectra and fermion mixings are studied in section 5. 
The results come up to phenomenological requirements. 
Whether we obtain the LMA-MSW solution or the SMA-MSW solution 
is controlled by the assignment of Abelian flavor charges. 
Final section is devoted to summary and discussion.

\vspace{10mm}

%%%%%  SECTION  2  %%%%%%%%%%%%%%%%%%%%%%%%%%%%%%%%%%%%%%%%%
\section{Discrete flavor symmetries}

In the string theory discrete symmetries stem from the symmetric 
structure of compact space. 
As a simple example of Calabi-Yau space let us consider 
the quintic hypersurface 
\be
  F(z) = \sum_{i=1}^{5} z_i^5 = 0 
\ee
in $CP^4$ with the homogeneous coordinates $z_i \ (i=1 \sim 5)$. 
On this Calabi-Yau space we have the discrete symmetry 
\be
  G = S_5 \, \sdpro ({\bf Z}_5)^5/{\bf Z}_5\,, 
\ee
where five ${\bf Z}_5$'s represent the phase transformations 
$z_i \rightarrow \alpha^{n_i} \,z_i$ with $\alpha^5 = 1$ and 
with integers $n_i$. 
Massless matter fields are classified according to charges of 
the discrete symmetry. 
Then the discrete symmetry plays a role of the flavor symmetry. 
String compactification on this hypersurface corresponds 
to the $3^5$ Gepner model\cite{Gepner}. 
In Gepner model the compact space is algebraically constructed in 
terms of a tensor product of discrete series of $N=2$ superconformal 
field theory with level $k_i$. 
When the trace anomaly condition 
\be
   \sum_{i=1}^{r} \frac{3k_i}{k_i + 2} = 9 
\ee
is satisfied, 
we have a complex 3-dimensional Calabi-Yau space with 
the discrete symmetry 
\be
   G = G_P \, \sdpro \prod_{i=1}^{r} {\bf Z}_{k_i + 2}/{\bf Z}_l \,, 
\ee
where $l$ is the l.c.m. of ${k_i + 2}\ (i=1 \sim r)$ and 
$G_P$ permutes the different variables of the same level $k_i$. 
The product $\prod_{i=1}^{r} {\bf Z}_{k_i + 2}$ is 
$R$ symmetry, while $G_P$ is non-$R$ symmetry. 
The Gepner model with $r=5$ corresponds to string compactification 
on weighted hypersurfaces in weighted $CP^4$
\cite{Greene}.

As seen from these examples of compact spaces, 
in the string theory there can appear various types of 
discrete $R$ symmetry and discrete non-$R$ symmetry as the 
flavor symmetry. 
Since the discrete symmertry group $G$ in the above examples 
contains semi-direct products of Abelian groups in addition to 
a permutation group, 
$G$ is non-Abelian as a whole. 
Thus it is natural to expect that the symmetry group is 
non-Abelian in general.

Let us first consider a compact space with an Abelian 
discrete symmetry. 
In string with discrete torsion 
the coordinates become non-commutative operators and 
are represented by a projective representation 
of the Abelian discrete symmetry\cite{torsion}. 
Hereafter the coordinates are referred to as quantum 
coordinates. 
We illustrate quantum coordinates with the above example. 
In the case of the quintic hypersurface in $CP^4$ 
quantum coordinates are described in terms of 
the projective representation of $({\bf Z}_5)^4$ 
and given by\cite{Beren} 
\bea
  \hat{z}_1 = z_1 \, P , \qquad \hat{z}_2 = z_2 \, Q, \qquad 
  \hat{z}_3 = z_3 \, P^{p_3} Q^{q_3},     \nonumber \\
  \hat{z}_4 = z_4 \, P^{p_4} Q^{q_4}, \qquad 
  \hat{z}_5 = z_5 \, P^{p_5} Q^{q_5}, \phantom{MM} 
\eea
where $p_i$ and $q_i$ $(i=3,4,5)$ are integers and 
\be
  P = {\rm diag}(1, \ \a, \ \a^2, \ \a^3, \ \a^4),  
   \qquad Q = \left(
      \begin{array}{ccccc}
        0 & 0 & 0 & 0 & 1 \\
        1 & 0 & 0 & 0 & 0 \\
        0 & 1 & 0 & 0 & 0 \\
        0 & 0 & 1 & 0 & 0 \\
        0 & 0 & 0 & 1 & 0 
      \end{array}
      \right) 
\ee
with $\alpha^5 = 1$. 
Since $P$ and $Q$ satisfy the relations 
\be
  P^5 = Q^5 = 1, \phantom{MM} 
  P \, Q = \a \, Q \, P, 
\ee
we have 
\be
  \hat{z}_1 \, \hat{z}_2 = \a \, \hat{z}_2 \, \hat{z}_1, \quad \cdots . 
\ee
The quantum fluctuation of coordinates is described in terms of 
the matrices $P$ and $Q$, 
which represent the non-commutativity in the compact space. 
When the quantum fluctuation is switched off, 
namely, in the string without discrete torsion, 
the matrix factors of $\hat{z}_i$ given by the products of 
$P$ and $Q$ disappear. 
In this case we are led to a compact space with commutative geometry 
(classical geometry)
\footnote{This situation is analogous to the appearance of the degree of 
freedom of spin in quantum mechanics. 
When we take the limit $\hbar \rightarrow 0$, 
coordinates commute with their conjugate momenta and also 
the degree of freedom of spin disappears.}. 
Massless matter fields in effective theory correspond to 
the degree of freedom of deformation of the compact space. 
In the above example the deformation is described in terms of 
polynomials of homogeneous coordinates $z_i$ with definite 
charges under the discrete group. 
This correspondence is expected to hold also in the case of 
non-commutative geometry. 
Since massless matter fields correspond to functions of 
the quantum coordinates $\hat{z}_i$, 
massless matter fields become matrix-valued. 
On the other hand, four-dimensional effective Lagrangian of 
the theory should belong to the center of 
the non-commutative algebra. 
This means that a new type of flavor symmetry arises 
in the compact space with non-commutative geometry.

We next consider an extension of the above argument 
to the compact space with a non-Abelian discrete symmetry. 
Based on the analogy to an Abelian discrete symmetry, we take 
the first ansatz that the quantum fluctuation of the coordinates 
is represented by a projective representation of 
the non-Abelian discrete symmetry. 
This ansatz implies that for coordinates we introduce 
the nontrivial commutation relation different from the usual 
canonical commutation relations\cite{Liecom,Qspace}. 
The second ansatz is that the degree of freedom of the deformation 
is expressed by the functions of quantum coordinates 
in the compact space with a certain non-Abelian discrete symmetry. 
Therefore, massless matter fields in effective theory are 
described in terms of quantum coordinates and become matrix-valued. 
Consequently, we have a new type of the flavor symmetry coming from 
the quantum fluctuation of coordinates.

In this paper we introduce ${\bf Z}_2  \, \sdpro {\bf Z}_4$ symmetry 
as the discrete non-Abelian flavor symmetry 
and consider its projective representation, 
which is motivated by phenomenological requirements 
as shown in section 4. 
The ${\bf Z}_2$ and ${\bf Z}_4$ groups are expressed as 
\bea
  {\bf Z}_2 & = & \{1, \ g_1 \}, \\
  {\bf Z}_4 & = & \{1, \ g_2, \ g_2^2, \ g_2^3 \}, 
\eea
respectively and we assume the relation 
\be
   g_1 g_2 g_1^{-1} = g_2^{-1}. 
\ee
The group ${\bf Z}_2  \, \sdpro {\bf Z}_4$ is nothing but 
the dihedral group $D_4$. 
The elements  $g_1$ and $g_2$ mean the reflection and 
$\pi/2$ rotation of a square, respectively. 
In view of the fact that $D_4 \subset SO(3) \subset SO(6)$ 
it is plausible that the dihedral group $D_4$ is contained 
in the discrete symmetry from six-dimensional string compactification. 
A vector representation of $D_4$ is given by 
\be
  \gamma_{g_1} = \left(
       \begin{array}{cc}
         0  & -i  \\
         i  &  0  
       \end{array}
       \right) = \sigma_2, \qquad 
  \gamma_{g_2} = \left(
       \begin{array}{cc}
         0  &  i  \\
         i  &  0  
       \end{array}
       \right) = i \sigma_1. 
\label{eqn:vecrep}
\ee
On the other hand, a projective representation of $D_4$ 
is given by 
\be
  \gamma_{g_1} = \left(
       \begin{array}{cc}
         0  & -i  \\
         i  &  0  
       \end{array}
       \right) = \sigma_2, \qquad 
  \gamma_{g_2} = \left(
       \begin{array}{cc}
         1  &  0  \\
         0  &  i  
       \end{array}
       \right) 
\label{eqn:prjrep}
\ee
and 
\be
   \gamma_{g_1} \gamma_{g_2} \gamma_{g_1}^{-1} = i \gamma_{g_2}^{-1}. 
\ee
We will use this projective representation of $D_4$ in section 4.

\vspace{10mm}

%%%%%  SECTION  3  %%%%%%%%%%%%%%%%%%%%%%%%%%%%%%%%%%%%%%%%%
\section{Minimal $SU(6) \times SU(2)_R$ model}

In this paper we take up minimal $SU(6) \times SU(2)_R$ 
string-inspired model, 
which has been studied in detail in 
Refs.\cite{Matsu1,Matsu2,Matsu3,CKM,MNS}. 
Here we briefly review the main points of the model. 
\begin{enumerate}
\item We choose $SU(6) \times SU(2)_R$ 
as the unification gauge symmetry at the string scale $M_S$, 
which can be derived from the perturbative heterotic 
superstring theory via the flux breaking\cite{Matsu4}. 

\item Matter chiral superfields consist of three families and one 
vector-like multiplet, i.e., 
%%%%%%%%%%%%%%%%%%%%%%%%%%%%%%%%%%%%%%%%%%%%%
\be
  3 \times {\bf 27}(\Phi_{1,2,3}) + 
        ({\bf 27}(\Phi_0)+\overline{\bf 27}({\bar \Phi})) 
\ee
%%%%%%%%%%%%%%%%%%%%%%%%%%%%%%%%%%%%%%%%%%%%%
in terms of $E_6$. 
Under $G_{gauge} = SU(6) \times SU(2)_R$, the superfields $\Phi$ in 
{\bf 27} of $E_6$ are decomposed into two groups as 
%%%%%%%%%%%%%%%%%%%%%%%%%%%%%%%%%%%%%%%%%%%%%%%%%%%%%%%%
\be
  \Phi({\bf 27})=\left\{
       \begin{array}{lll}
         \phi({\bf 15},{\bf 1})& : 
               & \quad \mbox{$Q,L,g,g^c,S$}, \\
          \psi({\bf 6}^*,{\bf 2}) & : 
               & \quad \mbox{$(U^c,D^c),(N^c,E^c),(H_u,H_d)$}, 
       \end{array}
       \right.
\label{27}
\ee
%%%%%%%%%%%%%%%%%%%%%%%%%%%%%%%%%%%%%%%%%%%%%%%%%%%%%%%%
where $g$, $g^c$ and $H_u$, $H_d$ represent colored Higgs and 
doublet Higgs superfields, respectively. 
$N^c$ is the right-handed neutrino superfield and 
$S$ is an $SO(10)$-singlet. 

\item Gauge invariant trilinear couplings in the superpotential 
$W$ become to be of the forms 
%%%%%%%%%%%%%%%%%%%%%%%%%%%%%%%%%%%%%%%%%%%%%%
\bea
    (\phi ({\bf 15},{\bf 1}))^3 & = & QQg + Qg^cL + g^cgS, \\
    \phi ({\bf 15},{\bf 1})(\psi ({\bf 6}^*,{\bf 2}))^2 & 
            = & QH_dD^c + QH_uU^c + LH_dE^c  + LH_uN^c 
                                            \nonumber \\ 
             {}& & \qquad   + SH_uH_d + 
                     gN^cD^c + gE^cU^c + g^cU^cD^c. 
\eea 
%%%%%%%%%%%%%%%%%%%%%%%%%%%%%%%%%%%%%%%%%%%%%%
\end{enumerate}

It has been found that this model contains phenomenologically 
attractive features. 
In the conventional GUT-type models, 
unless an adjoint or higher representation matter(Higgs) field 
develops a non-zero VEV, 
it is impossible that the large gauge symmetry is spontaneously 
broken down to the standard model gauge group $G_{SM}$ via 
Higgs mechanism. 
On the other hand, as explained above, in the present model 
matter fields consist only of ${\bf 27}$ and $\overline{\bf 27}$. 
The symmetry breaking of $G_{gauge} = SU(6) \times SU(2)_R$ 
down to $G_{SM}$ can take place via Higgs mechanism without 
adjoint or higher representation matter fields. 
In addition, $SU(6) \times SU(2)_R$ is one of the maximal 
subgroups of $E_6$. 
Further, it is noticeable that doublet Higgs and color-triplet 
Higgs fields belong to different irreducible representations 
of $G_{gauge}$. 
This situation is favorable to solve the triplet-doublet 
splitting problem.

As the flavor symmetry we introduce Abelian discrete $R$ 
and non-$R$ symmetries in the first place. 
Concretely, as for the $R$ symmetry we take $R$-parity and 
${\bf Z}_M$. 
We assign odd $R$-parity for $\Phi_{1,2,3}$ and even for $\Phi_0$ 
and $\overline{\Phi}$, respectively. 
Since ordinary Higgs doublets have even $R$-parity, 
they belong to $\Phi_0$. 
As for the non-$R$ symmetry we take 
${\bf Z}_N$. 
Assuming that $M$ and $N$ are relatively prime, 
we combine as 
\be
  {\bf Z}_M \times {\bf Z}_N = {\bf Z}_{MN}. 
\ee
In this case Grassmann number $\theta$ in superfield formalism 
has a charge $(-1, \ 0)$ under ${\bf Z}_M \times {\bf Z}_N$. 
The charge of $\theta$ under ${\bf Z}_{MN}$ is denoted as 
$q_{\theta}$. 
${\bf Z}_{MN}$-charges of matter superfields are denoted 
as $a_i$ and $b_i$, etc. as shown in Table 1.

\begin{table}
\caption{Assignment of ${\bf Z}_{MN}$-charge and $R$-parity 
          for matter superfields}
\label{table:1}
\bc
\begin{tabular}{|c|c|cc|} \hline 
    & \phantom{MM} $\Phi_i \ (i=1,2,3)\ $ & 
        \phantom{MM} $\Phi_0$ \phantom{MM} & 
              \phantom{MM} $\bar{\Phi}$ \phantom{MM} \\ \hline
$\phi({\bf 15, \ 1})$  & $(a_i, \ -)$ & $(a_0, \ +)$ 
                                & $(\bar{a}, \ +)$ \\
$\psi({\bf 6^*, \ 2})$ & $(b_i, \ -)$ & $(b_0, \ +)$ 
                                & $(\bar{b}, \ +)$ \\ \hline
\end{tabular}
\ec
\end{table}

In the superpotential there appear various types of 
non-renormalizable terms which respect both the gauge symmetry 
and the flavor symmetry. 
In $R$-parity even sector the superpotential contains the terms 
\be
  W_1 \sim M_S^3 \sum_{i=0}^r \l_i 
     \left( \frac{\phi_0 \bar{\phi}}{M_S^2} \right)^{n_i} 
         \left( \frac{\psi_0 \bar{\psi}}{M_S^2} \right)^{m_i}, 
\label{eqn:W1}
\ee
where $\l_i = \O(1)$ and the exponents are non-negative integers which 
satisfy the ${\bf Z}_{MN}$ symmetry condition 
\be
  n_i (a_0 + \bar{a}) + m_i (b_0 + \bar{b}) 
           - 2 q_{\theta} \equiv 0 \qquad {\rm mod} \ MN. 
\label{eqn:fc1}
\ee
Through the minimization of the scalar potential with 
the soft SUSY breaking mass terms characterized by 
the scale $\widetilde{m}_0 = \O(10^2)$GeV, 
matter fields develop non-zero VEV's. 
In Refs.\cite{Scale1} and \cite{Scale2} we have studied 
the minimum point of the scalar potential in detail. 
If and only if the relations 
\be
  n_i = (r-i) n_{r-1}, \qquad m_i = i m_1 \qquad (i=0 \sim r) 
\label{eqn:W1ex}
\ee
are satisfied, 
the gauge symmetry is spontaneously broken in two steps 
in a tolerable parameter region of the coefficients $\l_i$\cite{new}. 
Further, the scales of the gauge symmetry breaking are 
given by 
\bea
  |\langle \phi_0 \rangle| = |\langle \bar{\phi} \rangle| 
        & \sim & M_S \, \rho^{1/2(n_0-1)},  \nonumber  \\
  |\langle \psi_0 \rangle| = |\langle \bar{\psi} \rangle| 
        & \sim & M_S \, \rho^{n_{r-1}/2(n_0-1)m_1}, 
\label{eqn:scale}
\eea
where $\rho = \widetilde{m}_0/M_S \sim 10^{-16}$. 
The $D$-flat conditions require 
$|\langle \phi_0 \rangle| = |\langle \bar{\phi} \rangle|$ and 
$|\langle \psi_0 \rangle| = |\langle \bar{\psi} \rangle|$. 
Under the assumption $n_{r-1} > m_1$ 
we have 
\be
   |\langle \phi_0 \rangle| > |\langle \psi_0 \rangle|. 
\ee
Then the gauge symmetry is spontaneously broken 
at the scale $|\langle \phi_0({\bf 15, \ 1})\rangle |$ and 
subsequently at the scale 
$|\langle \psi_0({\bf 6^*, \ 2})\rangle |$. 
This yields the symmetry breakings 
\be
   SU(6) \times SU(2)_R 
     \buildrel \langle \phi_0 \rangle \over \longrightarrow 
             SU(4)_{\rm PS} \times SU(2)_L \times SU(2)_R  
     \buildrel \langle \psi_0 \rangle \over \longrightarrow 
     G_{SM}, 
\ee
where $SU(4)_{\rm PS}$ is the Pati-Salam $SU(4)$\cite{Pati}. 
Since the fields which develop non-zero VEV's are singlets under 
the remaining gauge symmetries, 
they are assigned as 
$\langle \phi_0({\bf 15, \ 1})\rangle = \langle S_0 \rangle $ and 
$\langle \psi_0({\bf 6^*, \ 2})\rangle = \langle N^c_0 \rangle $.
Below the scale $|\langle \phi_0 \rangle|$ Froggatt-Nielsen 
mechanism is at work for the non-renormalizable 
interactions\cite{F-N}.

Majorana masses for R-handed neutrinos are induced 
from the non-renormalizable terms 
\be
   M_S^{-1} \left( \frac {S_0 {\overline S}}{M_S^2} \right)^{\nu _{ij}} 
       (\psi_i {\bar \psi})(\psi_j {\bar \psi}) 
            \qquad (i,j = 1,2,3), 
\label{eqn:Majo}
\ee
where the exponents are given by 
\be
  (a_0 + \bar{a}) \nu_{ij} + b_i + b_j + 2\bar{b} 
      - 2 q_{\theta} \equiv 0 \qquad {\rm mod} \ MN. 
\label{eqn:fc2}
\ee
In fact, these terms lead to the Majorana mass terms 
\be
    {\cal N}_{ij} N^c_i N^c_j \sim x^{\nu_{ij}} 
        \left( \frac{\langle \bar{N^c} \rangle}{M_S} \right)^2 N^c_i N^c_j 
\label{eqn:MajoN}
\ee
in $M_S$ units, where we use the notation 
\be
   x = \frac {\langle S_0 \rangle \langle {\overline S} \rangle}{M_S^2}. 
\ee
From Eq.(\ref{eqn:scale}) we have 
\be
  x^{n_0-1} \sim \rho \sim 10^{-16}. 
\ee
Phenomenologically, it is desirable that the Majorana mass for the third 
generation is $\O(10^{10 \sim 12})$GeV. 
This scale is almost equal to the geometrically averaged value between 
$M_S$ and $M_W$, namely, 
\be
   x^{\nu_{33}} \left( \frac{\langle \bar{N^c} \rangle}{M_S} \right)^2 
          \sim \sqrt{\rho}. 
\ee
This is translated as 
\be
  \nu_{33} + \frac{n_{r-1}}{m_1} \sim \frac{n_0-1}{2}. 
\label{eqn:nu33}
\ee
Further, colored Higgs mass coming from 
\be
  \left( \frac {S_0 {\overline S}}{M_S^2} \right)^{\zeta _{00}} 
     S_0 g_0 g^c_0\,, 
\label{eqn:cHiggs}
\ee
is given by 
\be
  m_{g_0/g^c_0} = x^{\zeta_{00}} \, \langle S_0 \rangle. 
\ee
The ${\bf Z}_{MN}$ symmetry controls the exponent $\zeta_{00}$ as 
\be
  (a_0 + \bar{a}) \zeta_{00} + 3a_0 - 2 q_{\theta} \equiv 0 
                                              \qquad {\rm mod} \ MN. 
\label{eqn:fc3}
\ee
In order to guarantee the longevity of the proton, 
$\zeta_{00}$ should be sufficiently small compared to $n_0$. 
On the other hand, the $\mu$-term induced from 
\be
   \left( \frac {S_0 {\overline S}}{M_S^2} \right)^{\eta _{00}} 
       S_0 H_{u0} H_{d0}\,, 
\label{eqn:mu1}
\ee
is of the form 
\be
  \mu = x^{\eta_{00}} \, \langle S_0 \rangle. 
\ee
The exponent $\eta_{00}$ is determined by 
\be
  (a_0 + \bar{a}) \eta_{00} + a_0 + 2b_0 - 2 q_{\theta} \equiv 0 
                                              \qquad {\rm mod} \ MN. 
\label{eqn:fc4}
\ee
To be $\mu = \O(10^{2 \sim 3})$GeV, 
we need to obtain $\eta_{00} \sim n_0$ as a solution.

In order to find out a solution to the condition (\ref{eqn:nu33}) 
as well as $0 \leq \zeta_{00} \ll n_0$ and $\eta_{00} \sim n_0$, 
it is assumed that ${\bf Z}_{MN}$-charges of 
all matter superfields but $\bar{\phi}$ are even and 
that $a_0 + \bar{a} = -4$. 
Further, if $q_{\theta} = $ even and $q_{\theta} \ll MN$ and if 
$a_0 \equiv b_i + b_j \equiv MN - 2 \equiv 0 \ {\rm mod} \ 4$, 
we can expect that Eqs.(\ref{eqn:fc2}), (\ref{eqn:fc3}) and 
(\ref{eqn:fc4}) are reduced to 
\bea
  -4 \nu_{ij} + b_i + b_j + 2\bar{b} - 2 q_{\theta} & = & -MN,  \\
  -4 \zeta_{00} + 3a_0 - 2 q_{\theta} & = & 0,                  \\
  -4 \eta_{00} + a_0 + 2b_0 - 2 q_{\theta} & = & -2MN.      
\eea
Here we put $M=$odd and $N \equiv 2 \ {\rm mod} \ 4$ 
so as to render $q_{\theta}$ even. 
To be more specific, we choose a typical example 
\be
   M = 15, \qquad N = 14. 
\ee
In order to get $q_{\theta} \ll MN$ we take $|M - N| = 1$ 
and then $q_{\theta} = N$. 
Furthermore, for the sake of simplicity, 
$M$ and $N$ are chosen such that when 
decomposed into prime factors, 
all the prime factors are numbers with one figure. 
Thus we take the $R$ symmetry 
${\bf Z}_{15} = {\bf Z}_3 \times {\bf Z}_5$ 
and the non-$R$ symmetry 
${\bf Z}_{14} = {\bf Z}_2 \times {\bf Z}_7$. 
In this case we obtain $q_{\theta} = 14$ and $q_{\theta} \ll MN$. 
Further we put 
\be
  b_0 + \bar{b} = -49. 
\ee
Under these parametrizations Eq.(\ref{eqn:fc1}) becomes 
\be
   -4 n_i -49 m_i - 28 \equiv 0 \quad {\rm mod} \ 210. 
\label{eqn:nmi}
\ee
Since this equation allows the case 
\be
   -4 n_i -49 m_i - 28 = -210, 
\ee
we can not derive the relation (\ref{eqn:W1ex}). 
Then we need to forbid this case by introducing an additional 
selection rule. 
If the additional selection rule requires 
$m_i \equiv 0 \ {\rm mod} \ 4$, 
Eq.(\ref{eqn:nmi}) is rewritten as 
\be
  -4 n_i - 49 m_i - 28 = -420. 
\ee
This leads to 
\be
  (n_i, \ m_i) = (98, \ 0), \quad (49, \ 4), \quad (0, \ 8), 
\ee
which are in accord with the relation (\ref{eqn:W1ex}).

There appears similar situation in the $\mu$-term. 
In addition to the term (\ref{eqn:mu1}), 
the non-renormalizable term 
\be
   \left( \frac {S_0 {\overline S}}{M_S^2} \right)^{\eta' _{00}} 
      \left( \frac {N^c_0 {\overline N^c}}{M_S^2} \right)^{\xi} 
       S_0 H_{u0} H_{d0} 
\label{eqn:mu2}
\ee
is also allowed and leads to an additional $\mu$-term 
\be
  \mu' = x^{\eta'_{00} + \xi n_{r-1}/m_1} \, \langle S_0 \rangle. 
\ee
The exponents are determined by 
\be
  -4 \eta'_{00} -49 \xi + a_0 + 2b_0 - 28 \equiv 0 
                                              \quad {\rm mod} \ 210. 
\label{eqn:mud}
\ee
This allows the case 
\be
  -4 \eta'_{00} -49 \xi + a_0 + 2b_0 - 28 = -210. 
\ee
As a result we obtain an unrealistic solution 
$\mu' \gg \mu = \O(10^{2 \sim 3})$GeV. 
We also need to forbid this solution. 
If the additional selection rule requires 
$\xi \equiv 0 \ {\rm mod} \ 4$, 
then Eq.(\ref{eqn:mud}) is reduced to 
\be
  -4 \eta'_{00} -49 \xi + a_0 + 2b_0 - 28 = -420. 
\ee
In this case we can obtain $\mu, \ \mu' = \O(10^{2 \sim 3})$GeV. 
Thus we need an additional selection rule under which both the exponents 
$m_i$ and $\xi$ of $(\psi_0 \bar{\psi})$ in Eqs.(\ref{eqn:W1}) 
and (\ref{eqn:mu2}) should be multiples of 4.

\vspace{10mm}

%%%%%  SECTION  4  %%%%%%%%%%%%%%%%%%%%%%%%%%%%%%%%%%%%%%%%%
\section{Discrete non-Abelian symmetry}

As an additional flavor symmetry we introduce a discrete 
non-Abelian symmetry. 
It is postulated that due to the quantum fluctuation of coordinates 
this discrete non-Abelian symmetry is described 
in terms of the projective representation and 
that massless matter fields are matrix-valued. 
We denote matrix-valued charges of matter fields as 
$A_i$ and $B_i$, etc. as shown in Table 2. 
Since effective Lagrangian of the theory is the center of 
the non-commutative algebra, 
all the terms in the superpotential should be proportional 
to unit matrix provided that $\theta^2$ is neutral 
under the non-Abelian symmetry. 
Therefore, we have a new type of selection rule arising from 
the projective representation of non-Abelian symmetry.

\begin{table}
\caption{Assignment of matrix-valued charges}
\label{table:2}
\bc
\begin{tabular}{|c|c|cc|} \hline 
    & \phantom{M} $\Phi_i \ (i=1,2,3)\ $ & 
             \phantom{M} $\Phi_0$ \phantom{M} & 
                    \phantom{M} $\bar{\Phi}$ \phantom{M} \\ \hline
$\phi({\bf 15, \ 1})$   &  $A_i$  &  $A_0$  &  $\bar{A}$ \\
$\psi({\bf 6^*, \ 2})$  &  $B_i$  &  $B_0$  &  $\bar{B}$ \\ \hline
\end{tabular}
\ec
\end{table}

From the superpotential terms (\ref{eqn:W1}), (\ref{eqn:cHiggs}), 
(\ref{eqn:mu1}) and (\ref{eqn:mu2}) we have the non-Abelian 
symmetry conditions 
\bea
  \left(A_0 \bar{A} \right)^{n_i} 
                \left(B_0 \bar{B} \right)^{m_i} \propto 1,  \\
  \left( A_0 \bar{A} \right)^{\zeta _{00}} A_0^3 \propto 1,
\label{eqn:NAcHiggs}                                        \\
  \left( A_0 \bar{A} \right)^{\eta _{00}} 
                                      A_0 B_0^2 \propto 1,  \\
  \left( A_0 \bar{A} \right)^{\eta' _{00}} 
      \left(B_0 \bar{B} \right)^{\xi} A_0 B_0^2 \propto 1,  
\eea
respectively. 
When we put $A_0 \bar{A} = 1$, 
Eq.(\ref{eqn:NAcHiggs}) yields $A_0^3 \propto 1$. 
Accordingly, we take a simple choice $A_0 = \bar{A} = 1$. 
In this choice the above conditions are reduced to 
\be
  \left(B_0 \bar{B} \right)^{m_i}, \ 
  \left(B_0 \bar{B} \right)^{\xi}, \ B_0^2 \propto 1. 
\label{eqn:b02p}
\ee
As discussed in the previous section, we need the selection rule 
\be
m_i \equiv \xi \equiv 0 \qquad {\rm mod} \ 4, 
\ee
which are expressed as 
\be
   \left(B_0 \bar{B} \right)^4 \propto 1, \qquad 
    \left(B_0 \bar{B} \right)^2 \not \! \,\propto 1. 
\label{eqn:b02}
\ee

We turn to quark/lepton mass matrices. 
Mass matrix for up-type quarks comes from the term 
\be
  m_{ij} \left( \frac {S_0 {\overline S}}{M_S^2} \right)^{\mu_{ij}} 
       \, Q_i U^c_j H_{u0} 
\ee
with $m_{ij} = \O(1)$. 
Due to the Froggatt-Nielsen mechanism the mass matrix is given by 
\be
  {\cal M}_{ij} v_u = m_{ij} \, x^{\mu_{ij}} v_u. 
\ee
The exponent $\mu_{ij}$ is determined by 
\be
   -4\mu_{ij} + a_i + b_j + b_0 - 28 \equiv 0 
                                              \qquad {\rm mod} \ 210. 
\ee
In addition, the non-Abelian symmetry condition 
\be
  \left( A_0 \bar{A} \right)^{\mu_{ij}} A_i B_j B_0 = 
                        A_i B_j B_0 \propto 1 
\ee
is obtained. 
We here choose a solution under which this non-Abelian 
condition is satisfied irrespective of $i,\ j\ (i,j = 1,2,3)$. 
This choice is in line with the assumption that 
there is no texture-zero in quark/lepton mass matrices. 
Thus we put 
\be
  A_1 = A_2 = A_3, \qquad B_1 = B_2 = B_3 
\ee
and then obtain 
\be
   A_3 B_3 B_0 \propto 1. 
\label{eqn:a3b3b0}
\ee
If we took a different choice for matrix-valued charges $A_i$ 
and $B_i$ $(i=1,2,3)$, 
there could appear texture-zeros in quark/lepton mass matrices. 
Here we do not consider such a case.

In the down-quark sector mass matrix is given by\cite{Matsu1,Matsu2,Matsu3,CKM} 
\be
\begin{array}{r@{}l} 
   \vphantom{\bigg(}   &  \begin{array}{ccc} 
          \quad \,  g^c   &  \quad  D^c  &  
        \end{array}  \\ 
\widehat{{\cal M}}_d  = 
   \begin{array}{l} 
        g   \\  D  \\ 
   \end{array} 
     & 
\left( 
  \begin{array}{cc} 
    y_S {\cal Z}   &     y_N  {\cal M}  \\
      0     &  \rho_d  {\cal M} 
  \end{array} 
\right) 
\end{array} 
\label{eqn:Mhd}
\ee
in $M_S$ units, where $y_S = \langle S_0 \rangle /M_S$, 
$y_N = \langle N^c_0 \rangle /M_S$ and $\rho _d = v_d/M_S$. 
Since $g^c$ and $D^c$ have the same quantum number under 
the standard model gauge group, 
mixings occur between these fields. 
Consequently, mass matrix for down-type quarks becomes 
$6 \times 6$ matrix. 
The above $g\,$-$g^c$ submatrix coming from the term 
\be
  z_{ij} \left( \frac {S_0 {\overline S}}{M_S^2} \right)^{\zeta_{ij}} 
       \, S_0g_i g^c_j\,, 
\ee
is given by 
\be
   {\cal Z}_{ij} = z_{ij} \, x^{\zeta_{ij}} 
\ee
with $z_{ij}={\cal O}(1)$. 
The flavor symmetry requires the conditions 
\bea
  -4\zeta_{ij} + a_i + a_j + a_0 - 28 \equiv 0 \qquad {\rm mod} \ 210, \\
  \left( A_0 \bar{A} \right)^{\zeta_{ij}} A_i A_j A_0 =  
         A_3^2 \propto 1. \phantom{MMMM} 
\label{eqn:a32}
\eea

In the charged lepton sector mass matrix is of 
the form\cite{Matsu1,Matsu2,Matsu3,MNS} 
\be
\begin{array}{r@{}l} 
   \vphantom{\bigg(}   &  \begin{array}{ccc} 
          \quad   H_u^+   &  \quad  E^{c+}  &  
        \end{array}  \\ 
\widehat{{\cal M}}_l = 
   \begin{array}{l} 
        H_d^-  \\  L^-  \\ 
   \end{array} 
     & 
\left( 
  \begin{array}{cc} 
       y_S {\cal H}    &    0       \\
       y_N {\cal M}    &  \rho _d {\cal M} 
  \end{array} 
\right) 
\end{array} 
\label{eqn:Mhcl}
\ee
in $M_S$ units. 
Since $H_d$ and $L$ also have the same quantum number under 
the standard model gauge group, 
mixings occur between these fields. 
The above $H_d\,$-$H_u$ submatrix coming from 
\be
  h_{ij} \left( \frac {S_0 {\overline S}}{M_S^2} \right)^{\eta_{ij}} 
       \, S_0 H_{di} H_{uj}\,, 
\ee
is expressed as 
\be
   {\cal H}_{ij} = h_{ij} x^{\eta_{ij}} 
\ee
with $h_{ij}={\cal O}(1)$. 
From the flavor symmetry we have the conditions 
\bea
  -4\eta_{ij} + b_i + b_j + a_0 - 28 \equiv 0 \qquad {\rm mod} \ 210, \\
      \left( A_0 \bar{A} \right)^{\eta_{ij}} B_i B_j A_0 
          = B_3^2 \propto 1. \phantom{MMMM} 
\label{eqn:b32}
\eea

In the neutral sector there exist five types of matter fields 
$H_u^0$, $H_d^0$, $L^0$, $N^c$ and $S$. 
Then we have $15 \times 15$ mass matrix\cite{Matsu1,Matsu2,Matsu3,MNS} 
\be 
\begin{array}{r@{}l} 
   \vphantom{\bigg(}   &  \begin{array}{cccccc} 
          \quad \, H_u^0   &  \ \  H_d^0  &  \quad \ L^0  
                          &  \quad \ \ N^c   &  \quad \  S  &
        \end{array}  \\ 
\widehat{{\cal M}}_{NS} = 
   \begin{array}{l} 
        H_u^0  \\  H_d^0  \\  L^0  \\  N^c  \\  S  \\
   \end{array} 
     & 
\left( 
  \begin{array}{ccccc} 
       0     &  y_S {\cal H}     &   y_N {\cal M}^T     
                       &      0     &  \rho _d {\cal M}^T  \\
    y_S {\cal H}    &     0      &      0      
                       &      0     &  \rho _u {\cal M}^T  \\
    y_N {\cal M}    &     0     &      0      
                       &  \rho _u {\cal M} &       0       \\
       0     &     0     & \rho _u {\cal M}^T 
                       &      {\cal N}     &      {\cal T}^T      \\
   \rho _d {\cal M} & \rho _u {\cal M} &      0      
                       &      {\cal T}     &       {\cal S}       \\
  \end{array} 
\right) 
\end{array} 
\label{eqn:Mhn}
\ee
in $M_S$ units, where $\rho _u = v_u/M_S$. 
In this matrix the $6 \times 6$ submatrix 
\be
   \widehat{{\cal M}}_M = \left(
   \begin{array}{cc}
        {\cal N}    &   {\cal T}^T  \\
        {\cal T}    &    {\cal S}     
   \end{array}
   \right) 
\ee
play a role of Majorana mass matrix in the seesaw mechanism. 
The $3 \times 3$ submatrix ${\cal N}$ has already been given 
in Eq.(\ref{eqn:MajoN}). 
The flavor symmetry leads to the conditions 
\bea
  -4\nu_{ij} + b_i + b_j + 2\bar{b} - 28 \equiv 0 
                                \qquad {\rm mod} \ 210, \\
   \left( A_0 \bar{A} \right)^{\nu_{ij}} 
    (B_i \bar{B})(B_j \bar{B}) = (B_3 \bar{B})^2 \propto 1 \phantom{MM} 
\label{eqn:b3bb2}
\eea
The submatrix ${\cal S}$ induced from 
\be
   M_S^{-1} \left( \frac {S_0 {\overline S}}{M_S^2} \right)^{\sigma _{ij}} 
        (\phi_i {\bar \phi})(\phi_j {\bar \phi}), 
\ee
is expressed as 
\be
   {\cal S}_{ij} \sim x^{\sigma_{ij}} 
     \left( \frac{\langle \bar{S} \rangle}{M_S} \right)^2. 
\ee
The exponents are determined by 
\be
  -4\sigma_{ij} + a_i + a_j + 2\bar{a} - 28 \equiv 0 
                                \qquad {\rm mod} \ 210. 
\ee
The condition on matrix-valued charges 
\be
   \left( A_0 \bar{A} \right)^{\sigma_{ij}} 
    (A_i \bar{A})(A_j \bar{A}) = A_3^2 \propto 1 
\ee
is the same as Eq.(\ref{eqn:a32}). 
The submatrix ${\cal T}$ induced from 
\be
   M_S^{-1} \left( \frac {S_0 {\overline S}}{M_S^2} \right)^{\tau _{ij}} 
(\phi_i {\bar \phi})(\psi_j {\bar \psi}), 
\ee
is given by 
\be
   {\cal T}_{ij} \sim x^{\tau_{ij}} \frac{\langle \bar{S} \rangle 
                           \langle \bar{N^c} \rangle}{M_S^2}. 
\ee
The flavor symmetry yields the conditions 
\bea
  -4\tau_{ij} + a_i + b_j + \bar{a} + \bar{b} - 28 \equiv 0 
                                \qquad {\rm mod} \ 210, 
\label{eqn:tau}                                           \\
   \left( A_0 \bar{A} \right)^{\tau_{ij}} 
    (A_i \bar{A})(B_j \bar{B}) = (A_3 \bar{A})(B_3 \bar{B}) 
      \propto 1. \phantom{MM} 
\eea
However, since only $\bar{b}$ is taken as an odd integer, 
we have no solution to satisfy Eq.(\ref{eqn:tau}). 
This means the $3 \times 3$ matrix ${\cal T} = 0$.

Here we summarize the constraints on matrix-valued charges. 
First, we choose 
\be
  A_0 = \bar{A} = 1, \qquad A_1 = A_2 = A_3, \qquad B_1 = B_2 = B_3. 
\ee
From Eqs.(\ref{eqn:b02p}), (\ref{eqn:b02}), (\ref{eqn:a3b3b0}), 
(\ref{eqn:a32}), (\ref{eqn:b32}) and (\ref{eqn:b3bb2}) 
the conditions are put in order as 
\be
  A_3^2, \ B_3^2, \ B_0^2, \ A_3 B_3 B_0, \ 
  (B_3 \bar{B})^2, \ (B_0 \bar{B})^4 \propto 1, \quad 
     (B_0 \bar{B})^2 \not \! \,\propto 1. 
\label{eqn:NAcon}
\ee
If $[ \, B_3, \ \bar{B} \, ] = [ \, B_0, \ \bar{B} \, ] = 0$, 
these conditions are inconsistent. 
Consequently, it is necessary for us to introduce 
a non-Abelian symmetry as the flavor symmetry. 
The above conditions are realized provided that 
$B_3$, $\bar{B}$, $B_0$ and $A_3$ correspond to 
the elements $g_1$, $g_2$, $g_2^2$ and $g_1\,g_2^2$ in 
the dihedral group $D_4$ discussed in section 2. 
By taking the projective representation of 
$D_4$ 
\be
  A_3 = \sigma_1, \quad B_3 = \sigma_2, \quad 
  B_0 = \sigma_3, \quad 
  \bar{B} = \left(
       \begin{array}{cc}
         1  &  0  \\
         0  &  i  
       \end{array}
       \right), 
\label{eqn:g1g2}
\ee
where $\sigma_i$'s represent Pauli matrices, 
we obtain the relations 
\bea
   A_3^2 = B_3^2 = B_0^2 =1, \phantom{MM} \\
   A_3 B_3 B_0 = (B_3 \bar{B})^2 = i, \phantom{M}  \\
   (B_0 \bar{B})^2 = \sigma_3, \quad (B_0 \bar{B})^4 = 1. 
\eea
On the other hand, the conditions (\ref{eqn:NAcon}) 
are not satisfied in the case of the vector representation of 
$D_4$ given in Eq.(\ref{eqn:vecrep})
\footnote{If we take $4 \times 4$ matrices, 
we can obtain the vector representation of $D_4$ which 
satisfies the conditions (\ref{eqn:NAcon}). 
However, we are interested in the projective representation 
associated to the non-commutativity of coordinates.}. 

\vspace{10mm}

%%%%%  SECTION  5  %%%%%%%%%%%%%%%%%%%%%%%%%%%%%%%%%%%%%%%%%
\section{Mass spectra and fermion mixings}

In the previous section the dihedral group $D_4$ was introduced 
as an additional flavor symmetry. 
By using the dihedral flavor symmetry as well as Abelian $R$ and 
non-$R$ symmetries, 
we now proceed to study mass spectra and fermion mixings.

In section 3 we take the parametrization $M=15$, $N=14$ and 
\be
   a_0 + \bar{a} = -4, \qquad  b_0 + \bar{b} = -49. 
\ee
Further, it was assumed that ${\bf Z}_{210}$-charges of matter fields 
other than $\bar{b}$ are even and that $a_0 \equiv b_i + b_j \equiv 0 
\ {\rm mod} \ 4$. 
By virtue of the dihedral flavor symmetry offered here 
the superpotential (\ref{eqn:W1}) becomes 
\be
  W_1 \sim M_S^3 \left[ \l_0 
     \left( \frac{\phi_0 \bar{\phi}}{M_S^2} \right)^{98} 
       + \l_1 \left( \frac{\phi_0 \bar{\phi}}{M_S^2} \right)^{49} 
         \left( \frac{\psi_0 \bar{\psi}}{M_S^2} \right)^4 
       + \l_2 \left( \frac{\psi_0 \bar{\psi}}{M_S^2} \right)^8\right] 
\label{eqn:W1f}
\ee
and leads to 
\bea
  |\langle \phi_0 \rangle| = |\langle \bar{\phi} \rangle| 
        & \sim & M_S \, \rho^{1/194},  \nonumber  \\
  |\langle \psi_0 \rangle| = |\langle \bar{\psi} \rangle| 
        & \sim & M_S \, \rho^{49/776}. 
\label{eqn:scalef}
\eea
From the relation $x^{97} \sim \rho \sim 10^{-16}$ \ we obtain 
\be
  x^4 \simeq 0.23 \simeq \l. 
\label{eqn:x4}
\ee
As mentioned above, the gauge symmetry breaking takes place 
as 
\be
   SU(6) \times SU(2)_R 
     \buildrel \langle \phi_0 \rangle \over \longrightarrow 
             SU(4)_{\rm PS} \times SU(2)_L \times SU(2)_R  
     \buildrel \langle \psi_0 \rangle \over \longrightarrow 
     G_{SM}. 
\ee
At the first step of the symmetry breaking fields $Q_0$, $L_0$, 
${\overline Q}$, ${\overline L}$ and $(S_0 - {\overline S})/\sqrt{2}$ 
are absorbed by gauge fields. 
Through the subsequent symmetry breaking fields $U_0^c$, $E_0^c$, 
${\overline U}^c$, ${\overline E}^c$ and 
$(N_0^c - {\overline N}^c)/\sqrt{2}$ are absorbed.

The scale of colored Higgs mass is controlled by $\zeta_{00}$, 
which is determined by 
\be
  -4 \zeta_{00} + 3a_0 - 28 \equiv 0 \qquad {\rm mod} \ 210. 
\label{eqn:fc3f}
\ee
The $\mu$ is also controlled by $\eta_{00}$, 
which is determined by 
\be
  -4 \eta_{00} + a_0 + 2b_0 - 28 \equiv 0 \qquad {\rm mod} \ 210. 
\label{eqn:fc4f}
\ee
Therefore, if we put $420 \gg 3a_0 - 28 > 0$ and 
$-420 \ll a_0 + 2b_0 - 28 < 0$, then we have 
\bea
  -4 \zeta_{00} + 3a_0 - 28 & = & 0,     \\
  -4 \eta_{00} + a_0 + 2b_0 - 28 & = & -420,
\eea
where we used $a_0 \equiv 2b_0 \equiv 0 \ {\rm mod} \ 4$. 
In the case $a_0 = 12 \sim 20$, we have $\zeta_{00} = 2 \sim 8$ 
and then $m_{g_0/g^c_0} = \O(10^{17 \sim 18}){\rm GeV}$, 
which is consistent with the longevity of the proton. 
On the other hand, the parametrization 
$a_0 + 2b_0 = 0 \sim -20$ leads to a phenomenologically 
viable value of $\mu$, namely, 
$\mu = \O(10^{2 \sim 3}){\rm GeV}$.

We now return to quark/lepton mass matrices. 
Mass matrix for up-type quarks is given by 
${\cal M}_{ij} \sim x^{\mu_{ij}}$ with 
\be
  -4 \mu_{ij} + a_i + b_j + b_0 - 28 \equiv 0 \qquad {\rm mod} \ 210. 
\ee
In order to explain the experimental fact that top-quark mass is 
of $\O(v_u)$, 
we put 
\be
  a_3 + b_3 + b_0 - 28 = 0. 
\ee
Further, we take the parametrization 
\bea
  a_1 - a_3 = 48, & \qquad & a_2 - a_3 = 32, \nonumber \\
  b_1 - b_3 = 64, & \qquad & \, b_2 - b_3 = 32. 
\label{eqn:ab123}
\eea
As a result, the mass matrix ${\cal M}$ becomes 
\be
  {\cal M} \sim \left(
     \begin{array}{ccc}
       \l^7  &  \l^5  &  \l^3  \\
       \l^6  &  \l^4  &  \l^2  \\
       \l^4  &  \l^2  &   1  
     \end{array}
   \right), 
\ee
where we used Eq.(\ref{eqn:x4}). 
Eigenvalues of ${\cal M}$ yield up-type quark masses as 
\be
  (m_u, \ m_c, \ m_t) \sim (\l^7 v_u, \ \l^4 v_u, \ v_u). 
\ee

Down-type quark mass is derived from Eq.(\ref{eqn:Mhd}) with 
${\cal Z}_{ij} \sim x^{\zeta_{ij}}$, 
where 
\be
  -4\zeta_{ij} + a_i + a_j + a_0 - 28 \equiv 0 \qquad {\rm mod} \ 210. 
\ee
Then, under the parametrization (\ref{eqn:ab123}) we have 
\be
  {\cal Z} \sim \l^{\zeta} \left(
     \begin{array}{ccc}
       \l^6  &  \l^5  &  \l^3  \\
       \l^5  &  \l^4  &  \l^2  \\
       \l^3  &  \l^2  &   1  
     \end{array}
   \right),
\ee
where
\be
  \zeta = \frac{1}{16} ( a_3 - b_3 + a_0 - b_0) 
\ee
and we put $\zeta > 0$. 
Eigenstates of mass matrix (\ref{eqn:Mhd}) contain three heavy modes 
with their masses $\O(10^{17}){\rm GeV}$ and three light modes. 
When we choose 
\be
  \zeta \sim 2.5, 
\ee
light mode spectra turn out to be\cite{CKM} 
\be
   (m_d, \ m_s, \ m_b) \sim (\l^7 v_d, \ \l^6 v_d, \ \l^3 v_d). 
\ee
In addition, the above choice of $\zeta$ leads to the CKM matrix\cite{CKM} 
\be
V_{CKM} \sim \left(
  \begin{array}{ccc}
             1    &   \l    &   \l^5  \\
            \l    &    1    &   \l^2  \\
            \l^3  &   \l^2  &    1 
  \end{array}
  \right) 
\ee
at the string scale\cite{CKM}. 
It should be noticed that in the present model the element 
(1, 3) of $V_{CKM}$, i.e., $V_{ub}$ is suppressed compared 
with the element (3, 1), i.e., $V_{td}$.

In the charged-lepton sector the mass matrix is of the form 
(\ref{eqn:Mhcl}) with ${\cal H}_{ij} \sim x^{\eta_{ij}}$, 
where 
\be
  -4\eta_{ij} + b_i + b_j + a_0 - 28 \equiv 0 \quad {\rm mod} \ 210. 
\ee
The matrix ${\cal H}$ becomes 
\be
  {\cal H} \sim \l^{\eta} \left(
     \begin{array}{ccc}
       \l^8  &  \l^6  &  \l^4  \\
       \l^6  &  \l^4  &  \l^2  \\
       \l^4  &  \l^2  &   1  
     \end{array}
   \right), 
\ee
where 
\be
  \eta = \frac{1}{16} ( b_3 - a_3 + a_0 - b_0) \quad 
\ee
and we put $\eta \geq 0$. 
Eigenstates of mass matrix (\ref{eqn:Mhcl}) also contain three heavy modes 
with their masses $\O(10^{17}){\rm GeV}$ and three light modes. 
When we choose $\eta \sim 0$, 
light mode spectra are\cite{MNS} 
\be
  (m_e, \ m_{\mu}, \ m_{\tau}) \sim 
                (\l^7 v_d, \ \l^{4.5} v_d, \ \l^2 v_d). 
\ee
As seen soon later, this case corresponds to 
the LMA-MSW solution\cite{Atmos,Solar}. 
In the case $\eta \sim 2$, 
light modes become\cite{MNS} 
\be
  (m_e, \ m_{\mu}, \ m_{\tau}) \sim 
             (\l^{8.5} v_d, \ \l^5 v_d, \ \l^{2.5} v_d), 
\ee
which corresponds to the SMA-MSW solution\cite{Atmos,Solar}. 
In Table 3 we show interesting examples of assignment of 
${\bf Z}_{210}$-charges for matter superfields. 
From this Table we find 
\be
   \zeta_{00} = 2, \qquad \eta_{00} = 95, \qquad 
                     \zeta = 2.75, \qquad \eta = 0.25 
\ee
in the LMA-MSW solution and 
\be
   \zeta_{00} = 8, \qquad \eta_{00} = 95, \qquad 
                \zeta = 2.5, \qquad \eta = 2 \phantom{M} 
\ee
in the SMA-MSW solution.

\begin{table}
\caption{Assignment of ${\bf Z}_{210}$-charges for matter superfields}
\label{table:3}
\bc
\begin{tabular}{|c|c|c|} \hline 
    &     &   \\
    &   $\left(
\begin{array}{ccc}
 a_1 & a_2 & a_3 \\
 b_1 & b_2 & b_3 
\end{array} 
   \right)$   &  $\left(
\begin{array}{cc}
 a_0 & \bar{a} \\
 b_0 & \bar{b} 
\end{array} 
   \right)$   \\
   &       &     \\ \hline 
   &       &     \\
  $
\begin{array}{c}
{\rm LMA-MSW}  \\
{\rm solution} 
\end{array}
   $  &  $\left(
\begin{array}{ccc}
 78 & 62 & 30 \\
 74 & 42 & 10 
\end{array} 
   \right)$   & 
   $\left(
\begin{array}{cc}
  \phantom{-}12  & -16 \\
 -12  & -37 
\end{array} 
   \right)$       \\ 
   &       &      \\ \hline 
   &       &      \\
  $ 
\begin{array}{c}
{\rm SMA-MSW}  \\
{\rm solution} 
\end{array}
   $  & $\left(
\begin{array}{ccc}
 72 & 56 & 24 \\
 84 & 52 & 20 
\end{array} 
   \right)$ & 
   $\left(
\begin{array}{cc}
  \phantom{-}20  & -24 \\
 -16  & -33 
\end{array} 
   \right)$     \\ 
   &       &     \\ \hline 
\end{tabular}
\ec
\end{table}

In the neutral sector the mass matrix is given by Eq.(\ref{eqn:Mhn}) 
with ${\cal T}=0$. 
The mass matrix ${\cal S}$ is given by 
\be
    {\cal S}_{ij} \sim x^{\sigma_{ij}} 
           \left( \frac{\langle \bar{S} \rangle}{M_S} \right)^2 
            = x^{\sigma_{ij} + 1} 
\ee
with 
\be
  -4\sigma_{ij} + a_i + a_j + 2\bar{a} - 28 \equiv 0 
                                \quad {\rm mod} \ 210. 
\ee
In the LMA-, SMA-MSW solutions we obtain 
$\sigma_{33} = 0, \ 98$, respectively, which lead to 
\be
  {\cal S} \sim x \left(
     \begin{array}{ccc}
       \l^6  &  \l^5  &  \l^3  \\
       \l^5  &  \l^4  &  \l^2  \\
       \l^3  &  \l^2  &   1  
     \end{array}
   \right), \quad 
        x^2 \left(
     \begin{array}{ccc}
       \l^4  &  \l^3  &  \l  \\
       \l^3  &  \l^2  &   1  \\
       \l    &    1   &  \rho  
     \end{array}
   \right), 
\ee
respectively. 
From the above mass matrix eigenvalues of ${\cal S}$ are $\O(10^{14 \sim 18})$GeV 
and then sufficiently heavy compared to those of ${\cal N}$. 
Therefore, in seesaw mechanism an important role is played by 
the submatrix ${\cal N}$. 
The Majorana mass matrix ${\cal N}$ is given by 
\be
    {\cal N}_{ij} \sim x^{\nu_{ij}} 
           \left( \frac{\langle \bar{N^c} \rangle}{M_S} \right)^2 
         \sim x^{\nu_{ij} + 49/4} 
\ee
with 
\be
  -4\nu_{ij} + b_i + b_j + 2\bar{b} - 28 \equiv 0 
                                \quad {\rm mod} \ 210. 
\ee
From Table 3 this yields $\nu_{33} = 32, \ 39$ in the LMA-, 
SMA-MSW solutions, respectively. 
Then the Majorana mass for the third generation is obtained as 
\be
  m_{N^c_3} = \O(10^{11}){\rm GeV \ (LMA)}, \qquad 
              \O(10^{10}){\rm GeV \ (SMA)}. 
\ee
Thus we find the hierarchical Majorana mass matrix 
\be
  {\cal N} \sim (x^{44}, \ x^{51}) \times \left(
     \begin{array}{ccc}
       \l^8  &  \l^6  &  \l^4  \\
       \l^6  &  \l^4  &  \l^2  \\
       \l^4  &  \l^2  &   1  
     \end{array}
   \right)
\ee
in the LMA-, SMA-MSW solutions, respectively. 
It is worth noting that Dirac mass hierarchies in the neutrino sector 
cancel out with the Majorana sector in large part due to seesaw mechanism. 
Eigenstates of the $15 \times 15$ mass matrix (\ref{eqn:Mhn}) 
contain twelve heavy modes and three light modes. 
Light mode spectra turn out to be\cite{MNS} 
\bea
  (m_{\nu1}, \ m_{\nu2}, \ m_{\nu3}) & \sim & 
      \frac{v_u^2}{m_{N^c_3}} \times (\l^6, \ \l^5, \ \l^4), \\
  (m_{\nu1}, \ m_{\nu2}, \ m_{\nu3}) & \sim & 
      \frac{v_u^2}{m_{N^c_3}} \times (\l^9, \ \l^6, \ \l^5) 
\eea
in the LMA-, SMA-MSW solutions, respectively. 
Furthermore, the mixing angles in the MNS matrix become\cite{MNS} 
\be
  \tan \theta_{12} \sim \sqrt{\l}, \quad 
  \tan \theta_{23} \sim \sqrt{\l}, \quad 
  \tan \theta_{13} \sim \l, 
\ee
in the LMA-MSW solution and 
\be
  \tan \theta_{12} \sim \l^{1.5}, \quad 
  \tan \theta_{23} \sim \sqrt{\l}, \quad 
  \tan \theta_{13} \sim \l^2, 
\ee
in the SMA-MSW solution. 
It should be emphasized that the mixing angle $\tan \theta_{13}$ 
in the lepton sector becomes $\O(\l)$, $\O(\l^2)$ 
for the LMA-, SMA-MSW solutions, respectively. 
Whether we obtain the LMA-MSW solution or the SMA-MSW solution 
depends on the Abelian flavor charge assignment. 
Concretely, the solutions are governed by the parameter 
$\eta = (b_3 - a_3 + a_0 - b_0)/16$.

In the present model the massless sector at the string scale 
contains extra particles beyond the minimal supersymmetric 
standard model. 
In the course of the gauge symmetry breakings many particles 
become massive or are absorbed by gauge fields via Higgs 
mechanism at the intermediate energy scales. 
Therefore, integrating out these heavy modes we derive 
the low-energy effective theory in which large extra-particle 
mixings cause an apparant change of the Yukawa hierarchies 
for leptons and down-type quarks.

\vspace{10mm}

%%%%%  SECTION  6  %%%%%%%%%%%%%%%%%%%%%%%%%%%%%%%%%%%%%%%%%
\section{Summary and discussion}
String theory naturally provides discrete symmetries in which 
non-Abelian symmetry as well as Abelian $R$ and non-$R$ 
symmetries are contained as the flavor symmetry in the 
four-dimensional effective theory. 
Dihedral group $D_4$ introduced here is a possible example 
of non-Abelian symmetry expected in the string theory. 
In non-commutative compact space the coordinates become 
non-commutative operators and then are described in terms of 
the projective representation of the discrete symmetries. 
Explicitly, we take the projective representation of $D_4$. 
Since the deformation of the compact space is expressed 
by functions of the coordinates, 
massless matter fields in the effective theory are also 
described in terms of the projective representation and 
become matrix-valued. 
Thus we have a new type of selection rule coming from 
the projective representation of the discrete symmetry. 
This selection rule plays an important role in 
explaining the phenomenological 
fact that the scale of Majorana mass of R-handed neutrinos 
is almost equal to the geometrically averaged value between 
$M_S$ and $M_W$.

By using the dihedral flavor symmetry as well as Abelian $R$ 
and non-$R$ symmetries, 
we studied mass spectra and fermion mixings. 
Under an appropriate parametrization of the flavor charges 
our results come up to our expectations and are 
phenomenologically viable. 
The breaking scale of GUT-type gauge symmetry becomes 
$\O(10^{17 \sim 18})$GeV and the longevity of the proton 
is guaranteed. 
The scale of $\mu$ becomes $\O(10^{2 \sim 3})$GeV. 
Majorana mass of R-handed neutrinos for the third generation 
turns out to be $\O(10^{10 \sim 11})$GeV. 
An attractive account is also given of characteristic patterns 
of quark/lepton masses and mixing angles. 
Whether we obtain the LMA-MSW solution or the SMA-MSW solution 
depends on the Abelian charge parameter 
$\eta = (b_3 - a_3 + a_0 - b_0)/16$.

Finally, we touch upon the anomaly of the discrete symmetry 
${\bf Z}_{MN}$. 
If the ${\bf Z}_{MN}$ symmetry arises from certain gauge symmetries 
and if the anomaly cancellation does not occur via the 
Green-Schwartz mechanism\cite{G-S}, 
the ${\bf Z}_{MN}$ symmetry should be 
nonanomalous\cite{Ib-Ro,KMY}. 
Since the gauge symmetry at the string scale is assumed 
to be $SU(6) \times SU(2)_R$, 
the mixed anomaly conditions ${\bf Z}_{MN} \cdot (SU(6))^2$ 
and ${\bf Z}_{MN} \cdot (SU(2)_R)^2$ are imposed on 
${\bf Z}_{MN}$-charges of massless matter fields. 
In the present model the matter fields are $({\bf 15}, \ {\bf 1})$, 
$({\bf 6^*}, \ {\bf 2})$ and their conjugates under 
$SU(6) \times SU(2)_R$. 
Then we obtain the mixed anomaly conditions 
\be
  4 a_T + 2 b_T \equiv 18 q_{\theta}, \qquad 
   6 b_T \equiv 26 q_{\theta} \qquad {\rm mod} \ MN 
\ee
for $SU(6)$ and $SU(2)_R$, respectively, 
where 
\be
a_T = \sum_{i=0}^{3} a_i + \bar{a}, \qquad 
b_T = \sum_{i=0}^{3} b_i + \bar{b}. 
\ee
The present parametrization with $M=15$ and $N=14$ turns 
out to be inconsistent with the above anomaly conditions. 
The anomaly conditions yield stringent constraints on 
$M$, $N$ and ${\bf Z}_{MN}$-charge assignments for 
matter fields. 
Exploration into nonanomalous solutions with the discrete 
group ${\bf Z}_M(R) \times {\bf Z}_N({\rm non-}R)$ together 
with the dihedral group $D_4$ will be discussed elsewhere.

\vspace{10mm}

%%%%%  ACKNOWLEDGEMENTS  %%%%%%%%%%%%%%%%%%%%%%%%%%%%%%%%%%%
\section*{Acknowledgements}
Two of the authors (M. M. and T. M.) are supported in part by 
a Grant-in-Aid for Scientific Research, 
Ministry of Education, Culture, Sports, Science and Technology, 
Japan (No.12047226).

%%%%%  REFERENCES  %%%%%%%%%%%%%%%%%%%%%%%%%%%%%%%%%%%%%%%%%

%%%%  END  %%%%%%%%%%%%%%%%%%%%%%%%%%%%%%%%%%%%%%%%%%%%%%%%%

\end{document}